\documentclass[a4paper]{article}

\usepackage{ISCSLP2026}
\usepackage{ifthen}
\newboolean{blind}
\setboolean{blind}{false} 
\usepackage{multirow}
\usepackage{float}
\usepackage{booktabs}
\usepackage{comment}
\usepackage{makecell}
\usepackage{xcolor}
\usepackage{amsmath}
\usepackage{orcidlink}
\usepackage{booktabs}
\usepackage{CJKutf8}

\title{Mandarin Humorous Homophone Recognition and Disambiguation in Automatic Speech Recognition}

\name{
\ifthenelse{\boolean{blind}}{Anonymous to ISCSLP}{
Sicheng Jin\orcidlink{0009-0009-4492-8146}$^{1,*}$, 
Jinghao Chen\orcidlink{0009-0004-3707-0747}$^{2,*}$, 
Liuheng Zhou\orcidlink{0009-0009-0861-9071}$^2$,
Mostafa Shahin\orcidlink{0000-0002-1091-8531}$^2$, 
Beena Ahmed\orcidlink{0000-0002-1240-6572}$^2$, 
Aditya Joshi\orcidlink{0000-0003-2200-9703}$^1$
}
}

\address{
  \ifthenelse{\boolean{blind}}
  {Anonymous to ISCSLP}
  {
    $^1$School of Computer Science \& Engineering, UNSW Sydney, Australia \\
    $^2$School of Electrical Engineering \& Telecommunications, UNSW Sydney, Australia
  }
}

\email{
\ifthenelse{\boolean{blind}}
{Anonymous to ISCSLP}
{
\{z5317861,z5327748,z5271696,m.shahin,beena.ahmed,z3539958\}@unsw.edu.au
}
}

\begin{document}
\begin{CJK*}{UTF8}{gbsn}

\maketitle
\begingroup
\renewcommand{\thefootnote}{\fnsymbol{footnote}}
\footnotetext[1]{Equal contribution}
\endgroup

\begin{abstract}

Mandarin homophones remain a key challenge to improving automatic speech recognition (ASR) accuracy due to the amount of potential homophones. Mandarin speakers use this feature casually to convey emotions such as humour. Recent homophone-aware ASR studies have improved recognition accuracy, but intentional homophone twists in speech remain underexplored. In this paper, we identify patterns of homophone-based rhetorical wordplay in Mandarin, referred to as \emph{HumourPhone}, and propose an ASR Adapter for homophone and HumourPhone recovery. Experimental results show that the proposed approach improves the recall of recognising HumourPhone by over 5\% and achieves a 4.35\% drop in target-span character error rate for homophone correction compared to baseline. These results highlight the need for homophone-aware modelling of lexical ambiguity and rhetorical wordplay in Mandarin speech recognition.
\end{abstract}

\noindent\textbf{Index Terms}: speech recognition, Mandarin HumourPhone, computational paralinguistics


\section{Introduction}

Mandarin ASR systems and speech LLMs have achieved strong performance across a wide range of speech understanding tasks. However, they still struggle in homophone-rich scenarios, where the same or similar pronunciations map to semantically distinct written forms~\cite{sun2025context}. This challenge is amplified for rare, context-dependent, or deliberately playful forms, as models tend to output high-frequency lexical choices~\cite{fu2026pac}. Mandarin Chinese has a high density of homophones due to its tonal system and constrained syllable inventory; over 62.2\% of Chinese characters have been reported to exhibit homophonic features~\cite{sun2025context}. Speakers often exploit this ambiguity for rhetorical wordplay and humour, including homophonic puns, allegorical sayings, and online linguistic creativity~\cite{xu2024exploring,partington2009linguistic,alindra2026online}. Such wordplay may involve identical homophones with the same syllable and tone, or near homophones with small phonetic shifts~\cite{hiruncharoenvate2015algorithmically}. This makes Mandarin homophones challenging for speech models, as current speech LLMs still struggle to capture subtle prosodic and pitch cues that interact with syntax, semantics, and pragmatics~\cite{wang2025can}. In this work, \textit{HumourPhone} refers to utterances that evoke a familiar word or scenario through phonetic ambiguity and introduce an unexpected semantic twist~\cite{xu2026see,hong24homophone}.


Across languages, prior work has addressed homophones and phonetic ambiguity by explicitly incorporating pronunciation information. In Mandarin ASR, pronunciation-aware encodings, homophone-based label smoothing, and phonological speech attributes have been used to reduce character confusion and distinguish acoustically similar forms~\cite{zheng2020homophone,shen2023pronunciation,chen2026phonological}. Contextual biasing further uses homophone detectors to correct low-confidence outputs~\cite{yang2024contextual}. Related approaches include homophone-aware decoding for low-resource Cantonese ASR~\cite{2022Canto} and staged recognition and interpretation of spoken puns in English audio-language models~\cite{su2026words}. Together, these studies highlight the value of linking pronunciation with candidate written forms and contextual interpretation.

Recent Mandarin homophone studies provide further evidence for this pronunciation--orthography--context connection. Beyond phonological ambiguity, homophones have been studied as intentional devices in internet slang, memes, humour, and toxicity evasion~\cite{lin2026exploring,guo2025lost}. Existing methods therefore often combine phonological cues with higher-level interpretation, including pinyin-guided reasoning~\cite{ma2025reasoning}, multimodal chain-of-thought (CoT) verification for homophone and pun understanding~\cite{wang2024cieasr}, and template-based or sequence-to-sequence generation supported by human-engineered lexicons~\cite{su2025survey}. However, these methods generally rely on auxiliary textual, phonological, or lexicon-based information and are mainly designed for homophone disambiguation rather than speech-based humourphone, where the intended written form and humorous semantic twist must be inferred directly from audio.

In this work, we focus on speech-based Mandarin humourphone recognition and correction, aiming to identify intentional homophone-based wordplay in speech and restore its intended written form. Ordinary homophone correction is also reported as a secondary evaluation setting. We propose a task-conditioned ASR adaptation framework that detects potential humourphone spans, generates LLM-based candidates, and selects contextually appropriate corrections. Our contributions are: 1) a humourphone-oriented ASR adapter that also supports ordinary homophone correction; 2) a target-aware evaluation protocol for measuring recognition and correction within annotated regions; and 3) a humourphone dataset for evaluating speech models on homophone-based wordplay.

\begin{figure*}[t]
    \centering
    \includegraphics[width=0.9\textwidth]{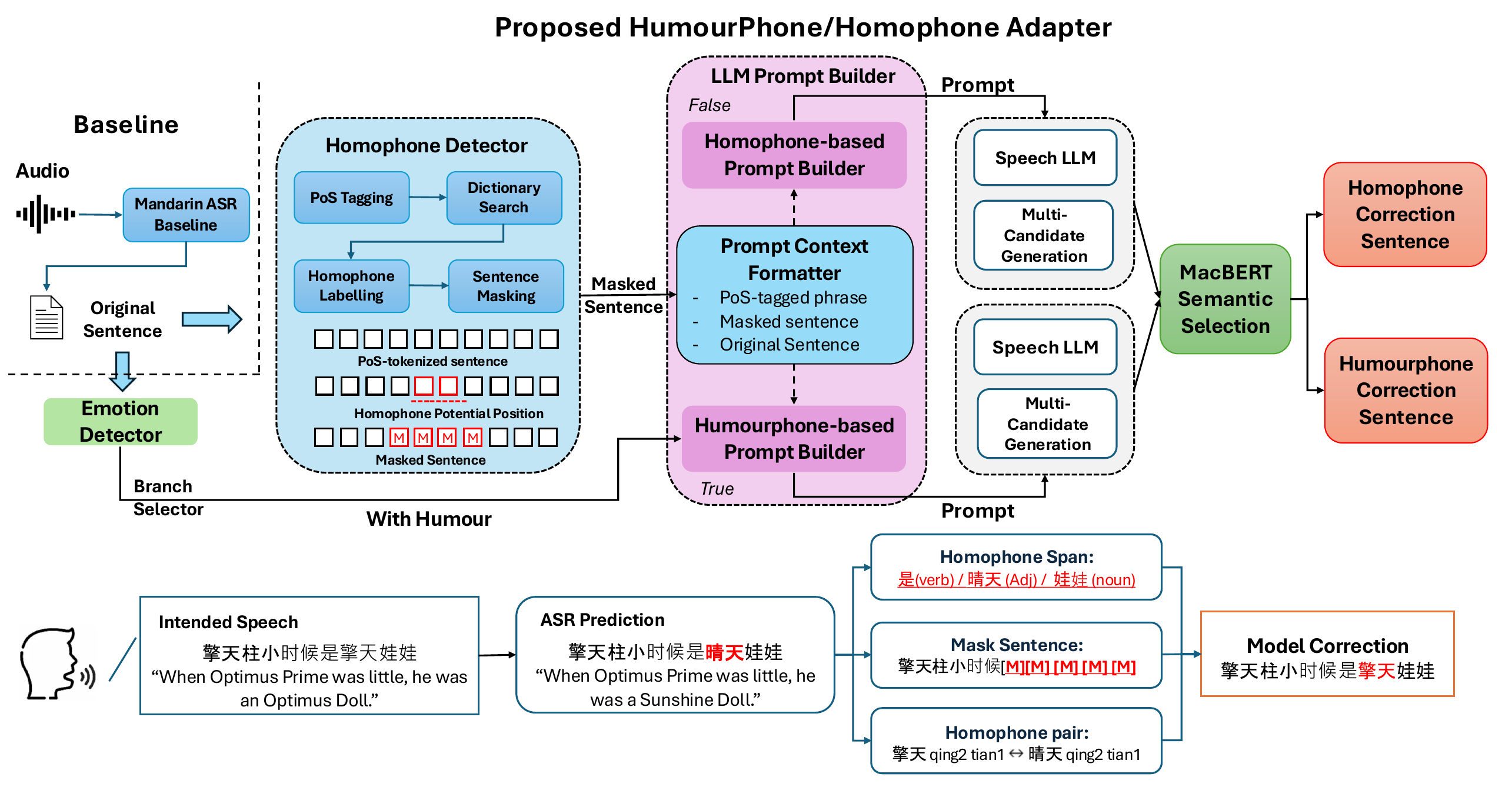}
    \caption{Workflow of the proposed adapter for potential-span-based homophone correction and humourphone recovery. In the bottom diagram, \textit{intended speech} refers to the input audio, \textit{ASR prediction} refers to the original sentence (i.e. baseline), while the three subsequent blocks are the output of the Homophone Detector.}  
    \label{fig:pipeline}
\end{figure*}

\section{Methodology}
Fig.~\ref{fig:pipeline} illustrates the proposed task-conditioned adapter for Mandarin homophone correction and humourphone recovery. We define homophones as words with similar or identical pronunciation but different meanings, and humourphones as homophones used intentionally for humorous effect. Given an utterance, the system first obtains an ASR transcription and routes it to either the homophone or humourphone branch using an emotion detector. Branch-specific contextual information is then converted into prompts for the speech LLM, which generates multiple candidate transcriptions from the original audio. A semantic selector chooses the final output while balancing correction accuracy and overcorrection risk.

\subsection{Humourphone Dataset Construction}
We collect Humourphone instances from internet sources, including text and images containing textual wordplay, and manually extract and normalise image text. As no publicly available Mandarin speech corpus specifically targets intentional humorous homophones, we synthesise Mandarin audio from the collected text using Microsoft Edge TTS~\cite{edgetts_misc}, resulting in 80 audio samples. The dataset therefore serves as a controlled evaluation set for humourphone recovery rather than a comprehensive representation of natural humorous speech. We manually remove samples with unreliable pronunciation, ambiguous polyphonic characters, inappropriate or non-humorous content, and rare derogatory expressions. The final set contains Full Homophones with identical pronunciation and Near Homophones with slight phonemic variation.

\subsection{Humourphone Pattern Guidelines}
\label{sec:humourphone_guidelines}
Based on the retained samples and literature on Mandarin homophonic wordplay~\cite{xu2026see,hong24homophone}, we identify recurring humourphone patterns to guide the emotion detector and the prompt builder in the humourphone branch. These cases typically involve a recognisable source expression, scenario, or discourse structure that is intentionally disrupted through identical or near-homophonic substitution. We summarise the observed patterns into six prompt-level guidelines: Phonetic Phrase Deviation, Phonetic Code-Switching, Rapid Q\&A and Punchline Structure, Anachronistic Semantic Clashes, Typographical and Visual Wordplay, and Anthropomorphic Attribution. These guidelines cover cases where homophones modify idioms or fixed expressions, imitate foreign words, letters, or numbers, create compact setup--punchline structures, or rely on contextual, visual, or anthropomorphic incongruity. Rather than serving as strict linguistic categories, these guidelines help distinguish intentional homophonic wordplay from ordinary ASR errors.

\subsection{Homophone and Humourphone Adapter}

The proposed adapter consists of a Homophone Detector, a branch-specific LLM prompt builder, and an output selector. Given an ASR transcription, the Homophone Detector performs PoS-aware span detection, while an Emotion Detector using an external API routes the utterance to either the homophone or humourphone branch.

We segment and PoS-tag the transcription using Jieba~\cite{jieba}. Suspicious lexical units and local combinations are identified by comparison with a Chinese word-frequency dictionary~\cite{cn_word_fre}. To account for segmentation uncertainty, we retain the top three suspicious spans and expand them with neighbouring PoS-tagged context before masking. The LLM is then prompted to generate two alternatives in addition to the original transcription, with modifications constrained primarily to the masked span to avoid unrestricted sentence-level rewriting. The homophone branch encourages homophone or near-homophone substitutions, while the humourphone branch additionally incorporates the pattern guidelines in Section~\ref{sec:humourphone_guidelines}.

For homophone correction, candidates are ranked using a MacBERT-based naturalness score~\cite{CN-Bert}. For a candidate transcription \(x=(x_1,\dots,x_N)\), we compute the masked pseudo-log-likelihood as
\begin{equation}
S(x)=\frac{1}{N}\sum_{i=1}^{N}
\log p_{\mathrm{MacBERT}}\left(x_i \mid x_{\mathrm{mask}}^{(i)}\right),
\label{eq:macbert_naturalness}
\end{equation}
where \(x_{\mathrm{mask}}^{(i)}\) denotes the transcription obtained by replacing token \(x_i\) with \texttt{[MASK]}. A higher \(S(x)\) indicates a more contextually natural candidate. The original transcription is retained when an alternative does not provide sufficient improvement, reducing overcorrection. In the humourphone branch, selection additionally considers consistency with the detected homophonic pattern while allowing the semantic incongruity characteristic of intentional wordplay.

\section{Experiments and Results}

\subsection{Data}
We construct Mandarin homophone evaluation sets from AISHELL-3 (AS-3)~\cite{AS3} and CommonVoice 25 Chinese (CV25-CN)~\cite{cv25}. We identify homophone targets using the ChineseHomophones dictionary~\cite{CN-homophone}. The final CN Dataset is formed by merging the subsets using unique pinyin-based homophone categories, and removing invalid targets such as person and location names. HumourPhone is excluded from this merged set due to its small scale and is evaluated separately for humour-oriented analysis. Table~\ref{tab:homophone_datasets} summarises the resulting datasets.

\begin{table}[ht]
  \caption{Statistics of the evaluation datasets. CN Dataset combines filtered AS-3 and CV25-CN, while HumourPhone is constructed for humour evaluation. Hom. Groups and Hom. Occ. denote homophone groups and occurrences.}
  \label{tab:homophone_datasets}
  \centering
  \scriptsize
  \setlength{\tabcolsep}{4pt}
  \renewcommand{\arraystretch}{1.1}
  \begin{tabular}{l c c c}
    \toprule
    \textbf{Dataset} & \textbf{Hom. Groups} & \textbf{Hom. Occ.} & \textbf{Hours} \\
    \midrule
    CN Dataset (with filter) & 1,303 & 85,706 & 125.73 \\
    HumourPhone (proposed)  & 80    & 80     & 0.08   \\
    \bottomrule
  \end{tabular}
\end{table}

\subsection{Experimental Setup}
We evaluate Mandarin homophone ASR on the merged CN Dataset in an inference-only setting. We compare Qwen3-ASR~\cite{qwen3} and Whisper-v3~\cite{whisper}, a prompt-only Qwen3-LLM baseline using a single generic homophone-aware instruction, and the proposed emotion-guided adapter. Unlike the baseline, the adapter prompt incorporates detected suspicious spans, masked local context, PoS information, and, for the humourphone branch, the pattern guidelines in Section~\ref{sec:humourphone_guidelines}. Qwen2.5-7B-Instruct~\cite{qwen2.5-llm} generates two alternatives plus the original transcription using deterministic decoding (\(T=0\), maximum 256 new tokens).

For frequency-based analysis, targets are labelled as \textit{common} or \textit{uncommon} using corpus frequency and a word-frequency dictionary~\cite{cn_word_fre}. The adapter uses Jieba~\cite{jieba} for span processing, Chinese MacBERT-large~\cite{CN-Bert} for naturalness-based selection, and GPT-5.5 via the OpenAI API~\cite{openai_gpt55} for routing. The emotion detector achieves 97.0\% and 76.0\% accuracy on Qwen3-ASR and Whisper-v3 HumourPhone transcripts, respectively, indicating sensitivity to upstream ASR outputs. All references and outputs are normalised to Simplified Chinese with numerals expanded into Mandarin written forms.

\subsection{Evaluation Metrics}
\label{sec:metric}
We evaluate Mandarin homophone ASR using target-span character error rate (T-CER), computed by character-level Levenshtein distance~\cite{Leven} over annotated target spans only. For correction, we report Recovery Rate (RR), the proportion of initially incorrect targets corrected by the pipeline, and Detection Rate (DetR), which measures whether an annotated target span overlaps with a predicted mask span.

For HumourPhone, we use T-CER and Mean Uncommon Recall (MUR), which measure recall over uncommon HumourPhone characters in the target span. We also report MacBERT-based Utterance-level Semantic Similarity (USS) and Target-span Semantic Similarity (TSS) to assess semantic consistency at full-utterance and target-span levels.

\subsection{Homophone Differentiation}

Table~\ref{tab:homophone_frequency} reports the effect of the proposed adapter on Whisper-v3 and Qwen3-ASR across common and uncommon homophone targets. We also evaluate a prompt-only speech-LLM baseline, Qwen3-LLM, but omit it from the table to focus on the adapter comparison. This baseline does not improve over Qwen3-ASR, with overall T-CER increasing slightly from 3.60\% to 3.70\%. This suggests that instruction-only speech-LLM decoding is insufficient for fine-grained homophone correction, as the model may struggle to locate suspicious homophone spans and can introduce incorrect substitutions.

With the proposed adapter, Whisper-v3 improves on both frequency groups. T-CER decreases from 6.43\% to 5.07\% on common forms, and more substantially from 24.74\% to 20.39\% on uncommon forms. The larger 4.35\% drop on uncommon homophones suggests that the adapter is particularly useful for correcting less frequent lexical choices. This is also reflected in the higher RR of Whisper-v3+Adapter, since Whisper-v3 produces more unstable lexical realisations and occasional phrase-level errors in Mandarin transcription, leaving more recoverable errors for span-level correction.

For Qwen3-ASR, the adapter reduces T-CER on uncommon homophones from 13.56\% to 12.85\%, indicating that span-level correction remains useful for less frequent lexical forms. In contrast, T-CER on common homophones increases from 2.19\% to 3.42\%. Since Qwen3-ASR already achieves a low error rate on common forms, there is limited room for further correction, and unnecessary substitutions can instead introduce errors into otherwise correct outputs. Qualitative inspection shows that many of the remaining Qwen3-ASR errors involve highly confusable lexical variants, such as the alternative written forms "部份" and "部分". Overall, this suggests that the adapter is most beneficial for difficult or uncommon homophones, while stronger ASR outputs require more conservative correction to avoid overcorrection.

\begin{table}[ht]
\centering
\footnotesize
\caption{Baseline (Base) and adapter-enhanced (Adapt.) ASR performance on the CN Dataset. T-CER, DetR, and RR denote target CER, detection rate, and recovery rate. \textit{Common}/\textit{uncommon} indicate high-/low-frequency reference forms. All values are percentages.}

\label{tab:homophone_frequency}
\renewcommand{\arraystretch}{1.1}
\setlength{\tabcolsep}{3pt}

\begin{tabular}{l l r c c c c}
\toprule
\textbf{Model}
& \textbf{Type}
& \textbf{Occ.}
& \multicolumn{2}{c}{\textbf{T-CER $\downarrow$}}
& \textbf{DetR $\uparrow$}
& \textbf{RR $\uparrow$} \\
\cmidrule(lr){4-5}
&
&
& \textbf{Base}
& \textbf{Adapt.}
&
& \\
\midrule
\multirow{2}{*}{Whisper-v3}
& Common
& 75,079
& 6.43
& \textbf{5.07}
& 37.22
& 7.18 \\

& Uncommon
& 10,627
& 24.74
& \textbf{20.39}
& 47.27
& 20.87 \\
\midrule

\multirow{2}{*}{Qwen3-ASR}
& Common
& 75,079
& \textbf{2.19}
& 3.42
& 44.42
& 2.79 \\

& Uncommon
& 10,627
& 13.56
& \textbf{12.85}
& 50.67
& 8.94 \\
\bottomrule
\end{tabular}
\end{table}






\begin{figure*}[t]
    \centering
    
    \includegraphics[width=1\linewidth]{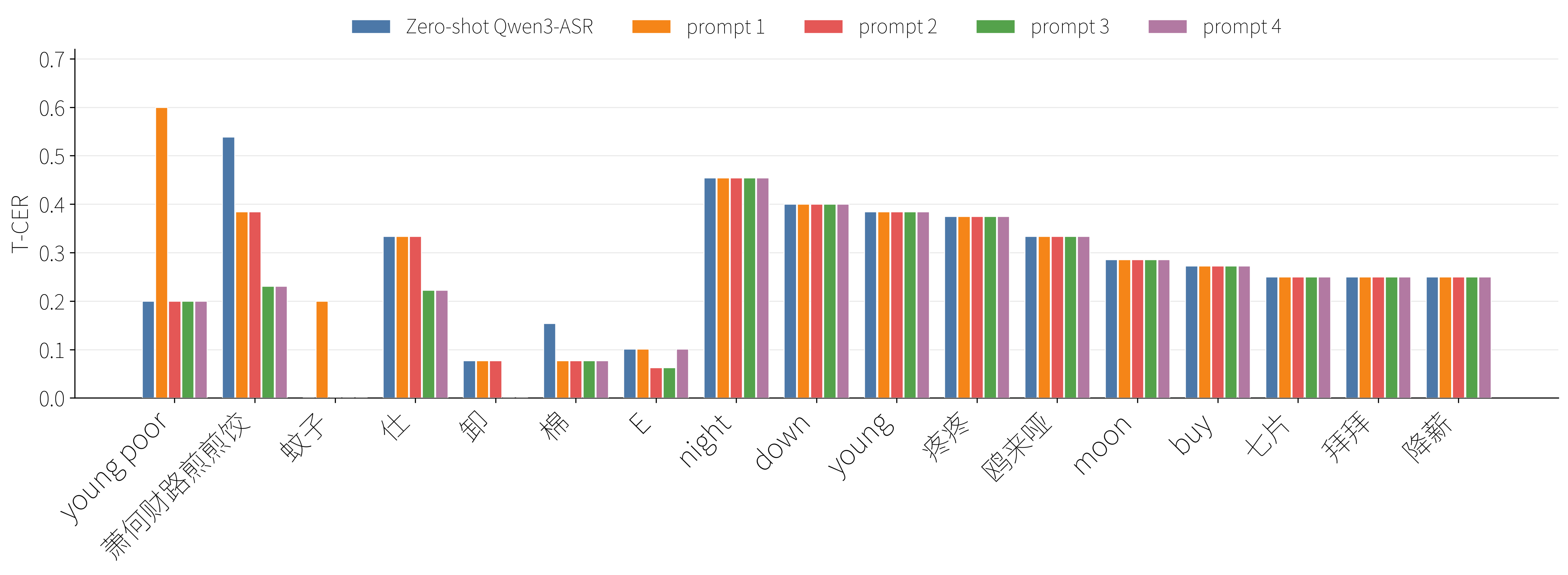}
    \caption{T-CER of different humourphone words transcribed by Qwen3-ASR across prompt variants. Some prompts reached T-CER=0 for certain words, hence the missing bars.}
    \label{fig:humourphone_prompt_cer}

    \vspace{1em} 
    
    \footnotesize
    \setlength{\tabcolsep}{3pt}
    \renewcommand{\arraystretch}{1.2}
    
    \captionof{table}{Comparison of predictions from ASR prompt variants for non-full-homophone, full-homophone, and overall humourphone cases. The USS and TSS are calculated from the MacBERT Semantic Selector. We compare the candidates produced by the Speech LLM with candidates produced by prompts constructed with the \textit{Homophone Detector}.}
    \label{tab:qwen-asr-humourphone-baseline}
    
    \resizebox{\textwidth}{!}{
        \begin{tabular}{l c c c c c c c c c c c c}
        \toprule
        & \multicolumn{4}{c}{\textbf{Near-homophone}} 
        & \multicolumn{4}{c}{\textbf{Full-homophone}} 
        & \multicolumn{4}{c}{\textbf{Overall}} \\
        \cmidrule(lr){2-5} \cmidrule(lr){6-9} \cmidrule(lr){10-13}
        \textbf{Method} 
        & \textbf{T-CER} $\downarrow$ & \textbf{MUR} $\uparrow$ & \textbf{USS} $\uparrow$ & \textbf{TSS} $\uparrow$
        & \textbf{T-CER} $\downarrow$ & \textbf{MUR} $\uparrow$ & \textbf{USS} $\uparrow$ & \textbf{TSS} $\uparrow$
        & \textbf{T-CER} $\downarrow$ & \textbf{MUR} $\uparrow$ & \textbf{USS} $\uparrow$ & \textbf{TSS} $\uparrow$ \\
        \midrule
        Qwen3-ASR & 11.61 & 0.6116 & 0.9902 & 0.9793 & 16.96 & 0.3095 & \textbf{0.9814} & 0.9614 & 15.21 & 0.4043 & 0.9854 & 0.9696 \\
        1-English-basic  & 13.15 & 0.6295 & 0.9924 & 0.9818 & 17.33 & 0.2810 & 0.9796 & 0.9596 & 15.97 & 0.3903 & 0.9855 & 0.9697 \\
        2-Chinese-basic  & 10.65 & 0.6295 & 0.9904 & 0.9798 & 16.50 & 0.3381 & \textbf{0.9814} & 0.9614 & 14.59 & 0.4295 & 0.9856 & 0.9697 \\
        3-Chinese-guidelines  & \textbf{9.69}  & \textbf{0.6473} & \textbf{0.9931} & \textbf{0.9823} & \textbf{15.92} & \textbf{0.3667} & 0.9812 & \textbf{0.9629} & \textbf{13.88} & \textbf{0.4547} & 0.9867 & \textbf{0.9718} \\
        4-Chinese-guidelines with example  & \textbf{9.69}  & \textbf{0.6473} & \textbf{0.9931} & \textbf{0.9823} & 16.15 & 0.3381 & \textbf{0.9814} & 0.9596 & 14.04 & 0.4351 & \textbf{0.9868} & 0.9700 \\
        \bottomrule
        \end{tabular}
    }
\end{figure*}

\subsection{HumourPhone Contextual Adaptation}
We further evaluate Qwen3-ASR~\cite{qwen3} on the constructed HumourPhone dataset, which contains humorous Mandarin utterances involving humour-bearing homophones. Unlike homophone evaluation, the target word may be intentionally ambiguous or semantically unexpected due to wordplay. This experiment examines whether ASR can recover such humour-bearing target spans under different levels of contextual guidance. As shown in Table~\ref{tab:qwen-asr-humourphone-baseline}, we compare a zero-shot Qwen3-ASR~\cite{qwen3} baseline with four prompt settings. In the zero-shot setting, the audio is transcribed without any task-specific instruction. For Prompts 1--4, we progressively modify the instructions used by the HumourPhone-based prompt builder, following the guidelines in Section~\ref{sec:humourphone_guidelines}. Initially, we provide the context only in English; for subsequent prompts, we change the instruction to Chinese. In prompt 3, we incorporate all the aforementioned guidelines, and we provide an example for prompt 4. We evaluate all settings using the metrics defined in Section~\ref{sec:metric}.

Contextual prompting can improve the model's ability to disambiguate intentional wordplay, although the effect depends on the prompt design. Prompt 1 slightly degrades performance relative to the zero-shot baseline, increasing the overall mean T-CER by 0.76\%, from 15.21\% to 15.97\%. However, with richer contextual background, this trend reverses starting with prompt 2. Prompt 3 achieves the strongest overall performance, reducing mean T-CER by 8.7\% relative, from 15.21\% to 13.88\%. The improvement is more pronounced for near-homophones, where T-CER decreases by 16.5\% relative, from 11.61\% to 9.69\%. Near-homophone TSS also increases by an absolute 0.003, reaching 0.9823. However, prompt 4 slightly worsened the overall performance. These results suggest that targeted contextual information generally helps the model select the intended, less frequent humourphone characters rather than defaulting to common homophones.

However, as illustrated in Figure~\ref{fig:humourphone_prompt_cer}, the gains remain limited, especially for full homophones. The difference in gains is not associated with whether the humourphone substitution is fully Chinese or bilingual. For example, the leftmost phrase "young poor" is a humourphone of the Chinese word "杨浦", which had a lower T-CER than the full Chinese pair "仕" and "事". Since full homophones share identical syllables and tones, the model cannot rely on phonetic deviation and must instead infer the intended humorous form from contextual and pragmatic cues. Therefore, prompting alone is insufficient for robust HumourPhone recovery. Future work should incorporate more HumourPhone examples and confidence-aware correction strategies to improve accuracy and reduce over-correction.

\section{Conclusion}
This paper proposed a task-conditioned ASR adapter for Mandarin homophone recognition and humourphone recovery. Results show that standard ASR systems favour common homophone forms, while simple LLM prompting provides limited gains. Adapter-based post-processing and targeted prompting show promising results for both ordinary homophone correction and humourphone recovery. The main limitations are the small synthetic HumourPhone dataset, lack of human evaluation, high LLM inference cost, and occasional overcorrection. Future work will expand natural-speech evaluation, incorporate human assessment, and reduce overcorrection.

\newpage


\bibliographystyle{IEEEtran}
\bibliography{mybib}
\end{CJK*}
\end{document}